\documentclass[twocolumn,aps,floats,showpacs,nofootinbib]{revtex4}

\usepackage{pifont,color,multirow,graphicx,mathrsfs,makeidx,epsfig,fancyhdr,fancybox,calc,
amsmath,amsfonts,amssymb,amsthm,latexsym,comment,appendix,ulem,lpic,tikz,caption}
\usepackage[latin1]{inputenc}
\usepackage{ragged2e} 

\usetikzlibrary{arrows}

\newcommand{\nc}{\newcommand}
\nc{\bb}{\begin{equation}} \nc{\ee}{\end{equation}}
\nc{\ug}{\; =\;} \nc{\tr}{\triangle} \nc{\rd}{{\rm d}} \nc{\R}{{\rm {I\!R}}}
\nc{\vs}{\vspace*} {\nc{\un}{1\!\!1} \nc{\0}{\vs*{0.5cm}}
\nc{\h}{\hspace*{0.5cm}} \nc{\vare}{\varepsilon}
\nc{\dis}{\displaystyle} \nc{\eee}{\end{document}}
\nc{\um}{\frac{1}{2}} \nc{\rdt}{\rd t} \nc{\pa}{\partial}
\nc{\munu}{{\mu\nu}} \nc{\C}{{\rm I\!\!\!C}}
\nc{\imp}{\mbox{\boldmath $p$}} \nc{\sbf}{\mbox{\boldmath $s$}} 
\nc{\kbf}{\mbox{\boldmath $k$}} \nc{\pibf}{\mbox{\boldmath$\pi$}}
\nc{\xbf}{\mbox{\boldmath $x$}} \nc{\nn}{\noindent} 

\nc{\tcb}{\textcolor{blue}} \nc{\tcr}{\textcolor{blue}}
\nc{\tcy}{\textcolor{yellow}} \nc{\tcg}{\textcolor{blue}}
\nc{\tcm}{\textcolor{magenta}} \nc{\tcc}{\textcolor{cyan}}
\nc{\tco}{\textcolor{orange}} \nc{\tcv}{\textcolor{violet}}

\nc{\pr}{^\prime} \nc{\ps}{/\!\!\!p} \nc{\Fs}{/\!\!\!F}

\begin{document}

\title{\mbox{Modified Lorentz transformations in deformed special relativity}\footnote{Work partially supported by INFN and MURST}}

\author{G. Salesi$^{\rm a,b}$\footnote{Corresponding author; e-mail: salesi@unibg.it}}

\author{M. Greselin$^{\rm a}$}

\author{L. Deleidi$^{\rm a}$}

\author{R.A. Peruzza$^{\rm a}$}

\affiliation{\ \\ \mbox{$^{\rm a}$Dipartimento di Ingegneria e
Scienze Applicate, Universit\`a di Bergamo,} viale Marconi 5,
Dalmine, Italy}

\affiliation{\mbox{$^{\rm b}$Istituto Nazionale di Fisica
Nucleare, Sezione di Milano, via Celoria 16, Milan, Italy}}

\

\

\begin{abstract}
\noindent We have extended a recent approach to Deformed Special Relativity based on deformed dispersion laws, entailing 
modified Lorentz transformations and, at the same time, noncommutative geometry and intrinsically discrete spacetime. 
In so doing we have obtained the explicit form of the modified Lorentz transformations for a special class of modified 
momentum-energy relations often found in literature and arising from quantum gravity and elementary particle physics. Actually, 
our theory looks as a very simple and natural extension of special relativity to include a momentum cut-off at the Planck 
scale. In particular, the new Lorentz transformations do imply that for high boost speed ($V \sim c$) the deformed 
Lorentz factor does not diverge as in ordinary relativity, but results to be upperly bounded by a large finite value 
of the order of the ratio between the Planck mass and the particle mass.  We have also predicted that a generic 
boost leaves unchanged Planck energy and momentum, which result invariant with respect to any reference frame.
Finally, through matrix deformation functions, we have extended our theory to more general cases with dispersion 
laws containing momentum-energy mixed terms.

\vspace*{0.1cm}

\nn \

\noindent

\end{abstract}

\maketitle


\newpage

\section{Introduction}

\noindent Quantum field theory and Standard Model, describing electromagnetic, weak, and strong interactions, 
are still to be unified with general relativity at the Planck scale, since classical and quantum theories entail very 
different predictions. In order to overcome this crucial trouble Lorentz violating (LV) theoretical approaches involve 
an upper energy-momentum scale and, due to the Heisenberg indeterminacy principle, also a fundamental 
spacetime scale.  Let us recall that an intrinsic length is directly correlated to the existence of a cut-off in the 
transferred momentum which avoids the occurrence of ``UV catastrophes'' in quantum field theories. Effective 
Lorentz covariance violations, which are to be investigated for very large particle momenta and energies, are found 
in different experimental and theoretical contexts \cite{All}: 
string and quantum gravity theories, grand-unification theories, non-covariant Standard Model Lagrangians containing 
Lorentz and CPT violating terms, foam-like quantum spacetimes, noncommutative geometries leading to a discrete 
spacetime at the Planck length or to a variable speed of light. In recent times Lorentz symmetry violations have also
been related to observation of ultra-high energy cosmic rays with energies beyond the Greisen-Zatsepin-Kuzmin 
cut-off \cite{GZK}. 

Some noticeable LV theories refer to the so called Deformed Special Relativity (DSR), endowed with deformed 
Lorentz symmetries, which, beyond the speed of light, involve a new elementary energy 
scale\footnote{Depending on the particular model, the energy scale can be the Planck mass $(10^{19}$\,GeV), 
or the GUT energy $(10^{15}$\,GeV), or the SUSY-breaking scale, 
or the superstring energy scale, and so on.}. A natural ``deformation'' of the standard dispersion law  $\,\varepsilon^2-p^2=m^2$ ($p \equiv |\imp|, \, \hbar = \rm{c} 
= 1$) can be put in most cases under the general form 
\bb
\varepsilon^2=p^2+m^2+p^2 f\left(\frac{p}{M}\right)
\label{uno}
\ee
where $M$ indicates a mass scale characterizing the Lorentz breakdown. Eq.\,(\ref{uno}) can be also rewritten as 
a series expansion with LV power terms ($a_i$ are dimensionless coefficients):
\bb
\varepsilon^2=p^2+m^2+p^2 \sum_{i=0}^\infty a_i\left(\frac{p}{M}\right)^i
\ee
\noindent 
In \cite{All,LVN}, by exploiting dispersion laws with the lower-order LV terms, quantitative bounds on Lorentz symmetry violation 
in the neutrino sector were obtained by analyzing laboratory data on neutron and pion decays. With analogous approach 
one of the authors (G.S.) in \cite{BALV} proposed a straightforward explanation for the cosmic matter-antimatter asymmetry 
based on a Lorentz-breakdown energy scale. A deformed dispersion law can also be written by means of ``form factors'' $f$ 
and $g$, expected to be different from 1 only for very high momenta:
\bb
g^2(p)\varepsilon^2 - f^2(p)p^2 = m^2  \label{Dispersion}
\ee
Correspondingly in recent past years it has been also proposed a momentum-dependent metric 
\bb \rd s^2 = g^{-2}(p)\rdt^2 -
f^{-2}(p)\rd l^2 \label{GRmetric}
\ee
which allows us to extend DSR to general relativity and to quantum gravity theories (cf. e.g., ``Rainbow Gravity'' 
theories\footnote{As in a rainbow or in a prism different wavelengths of light are differently refracted, analogously 
in deformed general relativity different de Broglie wavelengths move in separate trajectories and suffer different 
gravitational forces.}\cite{SM}.
A metric $g(p,x)$, $f(p,x)$ depending not only on spatial coordinates but also on particle energy ad momentum yields 
some interesting consequences and applications in cosmology and astrophysics \cite{BHE}.\\

\begin{figure}[!h]
\centering
\begin{tikzpicture}
\node [shape=rectangle,draw=black,rounded corners,align=center] (A) at (-3,0) {\footnotesize Modified dispersion law};
\node [shape=rectangle,draw=black,rounded corners,align=center](D) at (-1,-1) {\footnotesize Modified Lorentz transformations};
\node [shape=rectangle,draw=black,rounded corners,align=center] (E) at (1,0) {\footnotesize Deformed algebra};
\draw [<->](A) edge node[left] {$$} (D);
\draw [<->](A) edge node[right] {$$} (E);
\draw [<->](D) edge node[left] {$$} (E);
\end{tikzpicture}
\caption{\footnotesize Anyone of the three goals can be obtained from one of the other two.}
\label{Fig1}
\end{figure}
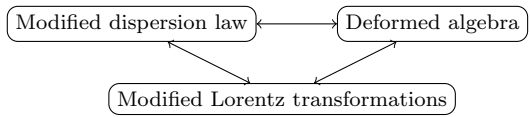

\nn In the present work our starting point is a suitable choice of modified dispersions laws from which we deduce 
modified Lorentz transformations (MLTs). Correspondingly, as shown in Section \ref{DLg}, any deformed
dispersion law yields a deformed Lorentz group. Modified dispersion law, modified Lorentz algebra and modified relativistic 
transformations are related to each other,
see Fig.\,\ref{Fig1}.

\noindent We shall parametrise deformed dispersion laws with a fundamental energy-momentum scale 
$\lambda$ which, when $\lambda$ goes to zero, recover the ordinary (SR) transformations. A flat spacetime 
metric will be throughout considered but, as abovesaid, the theory can be extended to include also curved spacetimes.
First we shall start with the most simple ``deformation function'' $F$
\bb
{F}^2\,(\varepsilon^2 \, - \, p^2 \, ) \, = \, m^2
\label{generalMDL}
\ee
where for $F$ it is initially chosen a $C$-number, depending on momentum and on ${\lambda}$; after we shall derive 
the explicit form of MLTs for more general deformation functions.

\section{Modified Lorentz Transformations \label{MLT}}

\vspace*{-0.5cm}
\nn Hereafter we shall choose deformation functions $F(p)$ (definite positive for any $p$ considered) depending on 
3-momentum $p$. Hereafter $p\equiv |\imp|$ whenever it appears without indices; the metric adopted for $g^\munu$ is $(+1; -1;-1; -1)$.
Moreover we have to assume the constraints
\bb
F(p=0)=1
\label{low_energy_limit}
\ee
[or, equivalently, $F(\varepsilon=m)=0$] which assures consistency with SR in the low energy limit, and 
\bb
F(\lambda=0)= 1
\label{Flambda}
\ee
which assures consistency with SR for small Lorentz symmetry violations, quantity $\lambda$ being the energy-momentum
cut-off parameter.

Let us now consider a wide class of physically meaningful deformation functions 
\bb
F^{(n)}=\left(1-\lambda^n p^n\right)^{-\frac{1}{n}} \qquad p\leq 1/\lambda\,, \ n \in \mathbb{N} - \{0\}
\label{Fn}
\ee 
(possible extensions of the present theory to non-integer $n$ can be studied as well).
This family of deformation functions entails the dispersion law 
\bb
\frac{\varepsilon^2-p^2}{(1-\lambda^n p^n)^\frac{2}{n}}=m^2
\label{disp_rel_n}
\ee
The previous modified dispersion law does involve a cut-off for the particle momentum 
\bb
p\leq \frac{1}{\lambda} 
\label{constraint}
\ee
We also assume that 
\bb
m \leq 1/\lambda
\label{masslambda}
\ee
that is, \,the rest mass must be smaller than the maximal mass $1/\lambda$ (e.g. than the Planck mass) or, equivalently, the 
Compton length $\hbar/mc$, the intrinsic quantum size of a particle, must be larger than the fundamental length $\lambda$  
(e.g. larger than the Planck length, assumed as the elementary step of the spacetime lattice in fundamental gauge theories).

For the above energy-momentum law when $p = 1/\lambda$ also $\vare=1/\lambda$, and viceversa. 
Moreover, as is easily proved, the 
value $1/\lambda$ for energy and momentum is a relativistic invariant: if energy 
and momentum are equal to $1/\lambda$ with respect to a given reference frame, they will be equal to $1/\lambda$ for any 
other inertial observer. Because of the existence of another ``absolute'' quantity besides the speed of light, DSR is also named 
as ``Doubly Special Relativity".

Some deformation functions analogous to the $n=1$ or $n=2$ case can be found in literature \cite{GAC,MS}, but with 
dispersion laws containing a pole in the energy rather than in the momentum as, e.g.,  the following
\bb
\frac{\varepsilon^2 - p^2}{(1-\lambda\varepsilon)^2}= m^2 \, \qquad F = (1-\lambda \varepsilon )^{-1} 
\label{A} 
\ee
\nn As is well-known, a generic operator $G$ transforms under a Lorentz boost along the $i$-axis according to the rule 
\bb
G^\prime = e^{-iN_i\psi}\,G\,e^{iN_i\psi}
\ee
where quantity $\psi$ represents the rapidity and $N_i$ is the boost operator i.e.\,the 4-rotation operator,
along the $i$-direction [$N_i\equiv M_{0i}$ for $i=1,2,3$, $M_{\mu\nu}$ being the angular momentum 
tensor). As an example, for the 4-momentum we have
\begin{eqnarray}
&&
p_\mu^\prime \ug p_\mu\,+ \,[p_\mu,iN_i\psi]\,+ \, \frac{1}{2!}\,[[p_\mu,iN_i\psi],iN_i\psi] \, \nonumber \\
&&
+ \frac{1}{3!}\,[[[p_\mu,iN_i\psi],iN_i\psi],iN_i\psi] \, + \, ... 
\label{rotazione}
\label{A8}
\end{eqnarray}
Inserting in the series the ordinary commutators holding in SR yields the standard Lorentz transformations 
for energy and momentum. Analogously, even in a DSR framework the MLTs can be derived exploiting the 
deformed operator algebra (cf.\,Eq.\,(\ref{boost_op_commutator})).  Usually the application of Eq.\,(\ref{rotazione}) 
results to be very long and expensive and we instead shall follow a more simple method. 
On considering (without loss of generality) a boost parallel to the $x$-axis, let us put our MLTs in the 
following form, often found in DSR literature \cite{Heuson},
\begin{eqnarray}
&&\left\{\begin{array}{ll}\varepsilon{\pr}=A\gamma(\vare - Vp_x) \\
p\pr_x=A\gamma(p_x - V\vare) \\
p\pr_y=Ap_y \\
p\pr_z=Ap_z 
\end{array}\right.
\label{DLTs1}
\end{eqnarray}
where $A$ is a real function of $p$ and quantity $A\gamma$ can be considered as a kind of ``deformed Lorentz factor''.

Indicating with $\Lambda$ the ordinary $4\times4$ LT matrix in special relativity
\[
\Lambda =
\begin {pmatrix}
\gamma & -\gamma V & 0 & 0 \\
-\gamma V & \gamma & 0 & 0 \\
0 & 0 & 1 & 0 \\
0 & 0 & 0 & 1 \\
\end {pmatrix}
\]
\nn we can concisely write MLTs (\ref{DLTs1}) as
\bb
p\pr_\mu = A\Lambda_{\mu\nu}p^\nu
\label{concise}
\ee
From (\ref{generalMDL}) and (\ref{concise}) we therefore obtain ($i=1,2,3$)
$$
F(p\pr)^2[{\vare\pr}^2 - {p\pr}^2] 
=F(p\pr)^2A^2\left[(\Lambda^{0\nu}p_\nu)^2 + \Lambda^{i\nu}p_\nu\Lambda_{i\rho}p^\rho\right] = 
$$
\bb
= F(p)^2[\vare^2 - p^2] = m^2
\label{FF}
\ee
Since (by definition of Lorentz $\Lambda$ matrix) quantity $(\Lambda^{0\nu}p_\nu)^2 +  \Lambda^{i\nu}p_\nu\Lambda_{i\rho}p^\rho$ 
turns out to be equal to $\vare^2-p^2$, Eq.\,(\ref{FF}) reduces to
\bb
F(p\pr)^2A^2 = F(p)^2 
\label{F2A2}
\ee
and then 
\bb
A = F(p\pr)^{-1}F(p)
\label{AFF}
\ee
Actually, we have chosen the positive solution of quadratic equation (\ref{F2A2}) in order that quantity $A(p)$ 
satisfies the small Lorentz violations limit $A (\lambda \to 0) = 1$. Starting from (\ref{AFF}) and taking into 
account that for (\ref{DLTs1})
\bb
p\pr = A[\gamma^2(p_x-V\varepsilon)^2+p_y^2+p^2_z]^\um
\ee
the deformation factor (and then the explicit expression of the MLTs) can be in
general obtained by solving an implicit equation in the unknown $A=A(p_x, p_y, p_z)$
\bb
\fbox{$A = F\left(A[\gamma^2(p_x-V\varepsilon)^2+p_y^2+p^2_z]^\um\right)^{-1}\!\!F(p)$}
\label{Impliciteq}
\ee
As is easy to show\footnote{C. Heuson: private communication.}, our MLTs own the usual group property for 
two subsequent boosts. If $\Lambda = \Lambda_1\Lambda_2$ is the ordinary Lorentz matrix for the total 
transformation, we shall obtain $p^{\prime}_\mu =  A\Lambda_\munu p^\nu$ where the global deformation factor 
turns out to be the product of the two deformation factors: $A = A_1A_2$.

\

\nn Let us now apply the ``resolvent'' equation \,(\ref{Impliciteq}) to the
family of deformed dispersion laws given by Eq.\,(\ref{Fn}).
The implicit equation (\ref{Impliciteq}) now writes
\bb
A = \frac{\left(1-\lambda^nA^n[\gamma^2(p_x-V\varepsilon)^2+p_y^2+p^2_z]^\frac{n}{2}\right)^{\frac{1}{n}}}{(1-\lambda^np^n)^{\frac{1}{n}}}
\ee
Thence, by raising both sides of the above equation to the $n^{\rm th}$ power
$$
A^n (1-\lambda^np^n)= 1-\lambda^nA^n[\gamma^2(p_x-V\varepsilon)^2+p_y^2+p^2_z]^\frac{n}{2}
$$
and collecting together all terms in $A^n$, we soon obtain the deformed Lorentz factor
\bb
A\gamma = \gamma\left\{1-\lambda^np^n+\lambda^n[\gamma^2(p_x-V\varepsilon)^2+p_y^2+p_z^2]^\frac{n}{2}\right\}^{-\frac{1}{n}}
\ee
Therefore our MLTs can be explicitly written as follows
\bb
\fbox{$\begin{array}{ll}
\vspace*{-0.3cm} \\
\left\{
\begin{array}{ll}
\displaystyle\varepsilon'
=\frac{\gamma\left(\varepsilon-Vp_x\right)}{\left\{1-\lambda^np^n+\lambda^n[\gamma^2(p_x-V\varepsilon)^2+p_y^2+p_z^2]^\frac{n}{2}\right\}^{\frac{1}{n}}} \\
\displaystyle p'_x
=\frac{\gamma\left(p_x-V\varepsilon\right)}{\left\{1-\lambda^np^n+\lambda^n[\gamma^2(p_x-V\varepsilon)^2+p_y^2+p_z^2]^\frac{n}{2}\right\}^{\frac{1}{n}}}\\
\displaystyle p'_y
=\frac{p_y}{\left\{1-\lambda^np^n+\lambda^n[\gamma^2(p_x-V\varepsilon)^2+p_y^2+p_z^2]^\frac{n}{2}\right\}^{\frac{1}{n}}}\\
\displaystyle 
p'_z=\frac{p_z}{\left\{1-\lambda^np^n+\lambda^n[\gamma^2(p_x-V\varepsilon)^2+p_y^2+p_z^2]^\frac{n}{2}\right\}^{\frac{1}{n}}}
\label{p_and_e_for_Fn}
\end{array}\right.
\\
\vspace*{-0.3cm} \\
\end{array}$}
\ee

\begin{figure}[!h]
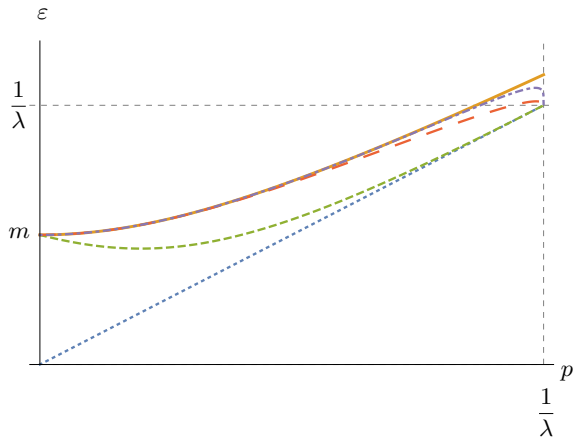

\begin{lpic}[l(7mm),r(4mm),t(4mm),b(7mm)]{epsilon_p(0.55)}
\lbl[t]{130,0;$p$}
\lbl[b]{125,-17;$\displaystyle \frac{1}{\lambda}$}
\lbl[l]{2,85;$\varepsilon$}
\lbl[l]{-5,32;$m$}
\lbl[l]{-5,63;$\displaystyle \frac{1}{\lambda}$}
\end{lpic}
\caption{\footnotesize Energy as a function of the momentum for different values of $n$: ordinary ($\lambda=0$) 
dispersion laws for massless (dotted line) and for massive particles (thick line); modified dispersion law 
($\lambda \neq 0$) for $n=1$ (dashed line), $n=4$ (large dashed line) and $n=10$ (dot-dashed line).}
\label{varepsilon(p)_1}
\end{figure}

\nn In Fig.\,\ref{varepsilon(p)_1} we see the deviation, growing for $p$ approaching the cut-off value 
$1/\lambda$, of the deformed dispersion law from the ordinary SR one, for any $n$.
Notice also that, as is easily proved exploiting the above MLTs, in the present LV theory the ordinary
velocity addition law still holds.

\nn Let us remark that, being from Eq.\,(\ref{FF})
\bb
{\vare\pr}^2 - {p\pr}^2 = A^2\,(\vare^2 - p^2) > 0
\label{ttssll}
\ee
the timelike nature of ordinary particles is conserved under a MLT: if $\vare>p$, also $\vare\pr>p\pr$.

\nn Furthermore, we could also extend the dispersion law (\ref{generalMDL}) to ``spacelike particles'' considering a 
deformation function which becomes negative, $F^2(p)<0$, in a given momentum interval. Actually,
for constraint (\ref{generalMDL}) a negative sign of $F^2$ would imply $\vare^2 - p^2 <0$.
Notice that also for $F^2<0$ it holds an equation identical to (\ref{F2A2}) 
$$
A^2 = {F(p\pr)}^{-2}F(p)^2
$$
where, obviously, multiplying the two negative quantities in the right side we obtain, as expected, a positive 
quantity for $A^2$. Therefore the momentum spacelike character would be conserved after a MLT: in fact 
for (\ref{ttssll}) if $\vare^2 - p^2 <0$ even ${\vare\pr}^2 - {p\pr}^2<0$. As an example, $F^2$ given by (\ref{Fn}) 
for $n=2$ becomes negative for $p>1/\lambda$:\footnote{For $n=2$ $\vare,p\,\lesseqqgtr\,1/\lambda$ implies $\vare\,\gtreqqless\,p$.} 
for such momenta the particle is spacelike, $\vare^2 - p^2 < 0$, ${\vare\pr}^2 - {p\pr}^2  < 0$. 

Finally, for massless particles the standard lightlike relation is recovered: we have $\vare^2 - p^2 = 0$, $\vare = p$, for any energy
and momentum with respect to any reference frame.

Summarizing, the sign of the difference between energy and momentum is preserved under a MLT, which conserves
the (subplanckian) timelike, or (transplanckian) spacelike, or photon-like nature of particles, respectively.

Besides massless luxons, a new, second kind of watershed between timelike and spacelike particles are ``Planckian particles'', i.e. particles with $\vare=p=1/\lambda$ and $0< m\leq1/\lambda$. As abovesaid, differently 
from photons which have $\vare=p$ but energy-momentum depending on the reference frame, Planckian particles 
transform in Planckian particles: also $\vare\pr=p\pr=1/\lambda$.

\

\noindent In \cite{CC} Carmona and Cortes, in order to explain the ``tritium beta decay anomaly'' occurring for very 
low energy of the order of few eV, argue that a plausible dispersion relation for neutrinos may be given by 
Eq.\,($\ref{disp_rel_n}$) with $n=1$:
\bb
\frac{\varepsilon^2 - p^2}{(1-\lambda p)^2} = m^2 \
\ee 
In this special case the deformed Lorentz factor turns out to be
\bb
A\gamma = \gamma\left\{1-\lambda p+\lambda[\gamma^2(p_x-V\varepsilon)^2+p_y^2+p_z^2]^\frac{1}{2}\right\}^{-1}
\ee
Actually, as it can be seen in Fig.\,\ref{varepsilon(p)_1} (dashed line), differently from any $n \neq 1$ case, for $n=1$ 
the dispersion law plot departs from the ordinary one even for small momentum in the low energy sector.

\

\begin{figure}[!h]
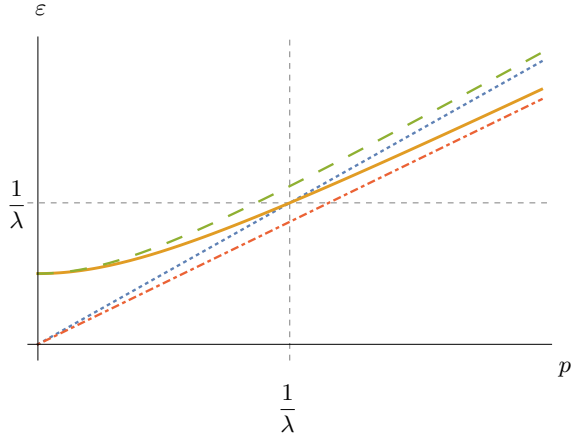

\centering
\begin{lpic}[l(7mm),r(4mm),t(4mm),b(7mm)]{epsilon_p_lambda(0.55)}
\lbl[t]{130,0;$p$}
\lbl[b]{63,-17;$\displaystyle \frac{1}{\lambda}$}
\lbl[l]{2,85;$\varepsilon$}
\lbl[l]{-5,38;$\displaystyle \frac{1}{\lambda}$}
\end{lpic}
\caption{\footnotesize Energy plotted as a function of the momentum for $n=2$ case: ordinary ($\lambda=0$) dispersion laws for massless (dotted line) and for massive 
particles (dashed line); modified dispersion law ($\lambda \neq 0$) (thick line) and its asymptotic behavior (dot-dashed line).}
\label{varepsilon(p)_2}
\end{figure}

\nn 
Another important deformation function, found in effective QFTs at the Planck scale and in basic 
string theories, is the one obtained taking $n=2$ in (\ref{Fn}) 
\bb
\frac{\varepsilon^2-p^2}{1-\lambda^2 p^2}=m^2 
\label{C9}
\ee
\nn This momentum-energy relation can be equivalently rewritten in the following form
\bb
\varepsilon^2-p^2=m^2(1-\lambda^2 p^2)
\label{transverse_mass}
\ee
where it appears a kind of ``variable mass'' containing the term, often called ``transverse mass'', $-\lambda^2 p^2m^2$, 
which is a (negative) momentum function quantity. Such transverse mass is in a sense analogous to the dynamical mass
of photons in superconductors due to the presence of a discrete atomic lattice as propagation medium (see comments
to Eq.\,(\ref{imp_operator}) at the end of the Section \ref{DLg}). Eq.\,(\ref{transverse_mass}) yields the particle energy as a momentum function, 
plotted in Fig.\,$\ref{varepsilon(p)_2}$:
\bb
\displaystyle
\varepsilon(p)=\sqrt{p^2(1-\lambda^2m^2)+m^2}
\label{n=2law}
\ee
Notice that for $p<1/\lambda$ the deformed dispersion plot is always contained between the ordinary SR dispersion 
plot and the massless dispersion plot, crossing the last one in $p=1/\lambda$. For $p \longrightarrow 1/\lambda$, a 
first order approximation of the energy is
\bb
\varepsilon(p) \sim \frac{1}{\lambda} +\alpha^2\left(p-\frac{1}{\lambda}\right)
\ee
where 
\bb
\alpha \equiv (1-\lambda^2m^2)^{\frac{1}{2}}
\label{alpha}
\ee
In Fig.\,$\ref{varepsilon(p)_2}$ we have quoted (dot-dashed line) also the asymptotic trend $\varepsilon(p) \sim \alpha p$ 
for $p \longrightarrow \infty$.

\nn Eqs.\,(\ref{p_and_e_for_Fn}) now reduce to 
\bb
\left\{
\begin{array}{ll}
\displaystyle\varepsilon\pr
=\frac{\gamma\left(\varepsilon-Vp_x\right)}{\left\{1-\lambda^2p^2+\lambda^2[\gamma^2(p_x-V\varepsilon)^2+p_y^2+p_z^2]\right\}^{\frac{1}{2}}}\\
 \displaystyle p\pr_x
=\frac{\gamma(p_x-V\varepsilon)}{\left\{1-\lambda^2p^2+\lambda^2[\gamma^2(p_x-V\varepsilon)^2+p_y^2+p_z^2]\right\}^{\frac{1}{2}}}\\
  \displaystyle p\pr_y
=\frac{p_y}{\left\{1-\lambda^2p^2+\lambda^2[\gamma^2(p_x-V\varepsilon)^2+p_y^2+p_z^2]\right\}^{\frac{1}{2}}}\\
  \displaystyle p\pr_z
=\frac{p_z}{\left\{1-\lambda^2p^2+\lambda^2[\gamma^2(p_x-V\varepsilon)^2+p_y^2+p_z^2]\right\}^{\frac{1}{2}}}
\end{array}
\right.
\ee
\nn with deformed Lorentz factor
\bb
A\gamma = \gamma\left\{1-\lambda^2p^2+\lambda^2[\gamma^2(p_x-V\varepsilon)^2+p_y^2+p_z^2]\right\}^{-\um}
\ee

\nn Let us recall that the previous result can be obtained also by inserting in Eq.\,(\ref{A8}) the deformed Lorentz commutators for the case $n=2$ quoted in Eqs\,(\ref{Nvare}) and (\ref{Np}).

\nn Some interesting properties of $A$ are the following:
\begin{itemize}
\item[a)] for small particle momentum ($p \sim 0$) $A\gamma$ varies from 1 for small boost, $V \sim 0$, to $1/m\lambda$ 
for maximal boost, $V \sim 1$
\bb
\framebox(100,30){$
\displaystyle \lim_{V \to 1} A\left(0,\lambda\right) \gamma = \frac{1}{m \lambda}
$}
\label{Agamma_limit}
\ee
\item[b)] for large particle momentum ($\displaystyle p \sim 1/\lambda$) $A\gamma$ ranges from 1 ($V \sim 0$) to infinity ($V \sim 1$)
\end{itemize}

\begin{figure}[!h]
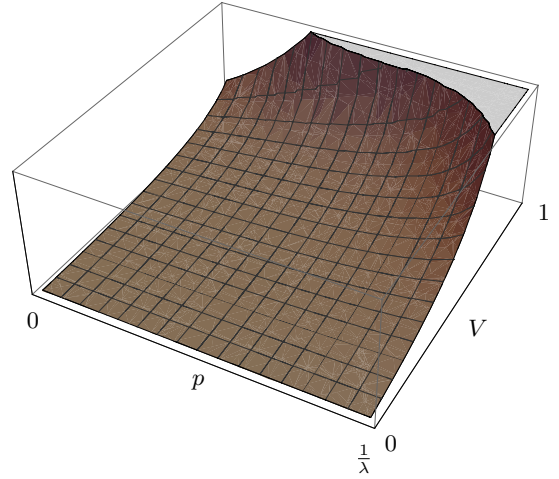

\centering
\begin{lpic}[l(7mm),r(4mm),t(4mm),b(7mm)]{red-Agamma3D(0.55)}
\lbl[b]{45,10;$p$}
\lbl[b]{5,25;$0$}
\lbl[b]{85,-10;$\frac{1}{\lambda}$}
\lbl[r]{115,25;$V$}
\lbl[r]{130,53;$1$}
\lbl[r]{93,-3;$0$}
\end{lpic}
\caption{\footnotesize Deformed Lorentz factor $A\gamma$ plotted as a function of $p$ and $V$ for $n=2$.}
\label{Agamma}
\end{figure}

\noindent Quite interestingly, it can be shown that both the above properties hold for a generic $n$ ($\neq 2$) as well.
The finite limit (\ref{Agamma_limit}) obtained for the modified Lorentz factor could involve typical relativistic effects 
(as e.g.\,length contraction, time dilation, mass increase, and so on) to be not infinite anymore but upperly bounded also 
for quasi-luminal boosts with respect to the rest frame.\\

\nn By applying a $-V$ boost along the $x$-axis to the particle rest frame, where $\varepsilon_0=m, \,\imp_0=0$, we pass to a 
laboratory frame where the particle acquires a speed $V$
\begin{eqnarray}
&&
\left\{\begin{array}{ll}
\varepsilon=A \gamma (\varepsilon_0+Vp_{0x})= A \gamma m\\
p_x(=p) = A \gamma (p_{0x}+V\varepsilon_0)= A \gamma V m\\
p_y = 0\\
p_z = 0
\label{DSR_epsilon_p}
\end{array}
\right.
\end{eqnarray}
Since in the rest frame
\bb
A=(1+\lambda^2\gamma^2V^2m^2)^{-\frac{1}{2}}
\label{A_n=2}
\ee
from Eqs.\,(\ref{DSR_epsilon_p}) we suddenly obtain $\varepsilon(V)$ and $p(V)$ [$\alpha$ is the constant 
previously defined in (\ref{alpha})]: 

\begin{eqnarray}
&&
\displaystyle \varepsilon= \frac{m}{\sqrt{1-\alpha^2 V^2}}
\label{E(V)} \\
&&
\displaystyle p= \frac{mV}{\sqrt{1-\alpha^2 V^2}}
\label{p(V)}
\end{eqnarray}

\

\nn The particle energy, Eq.\,(\ref{E(V)}), starts from the minimum value $m$ for $V=0$ and then approaches the Planck value 
$1/\lambda$ for $V \longrightarrow 1$. Because of (\ref{A_n=2}) DSR corrections to relativistic phenomena become not negligible 
for boost speed $V$ not too small compared to $1/\sqrt{1+m^2\lambda^2}$. 

\

\begin{figure}[!h]
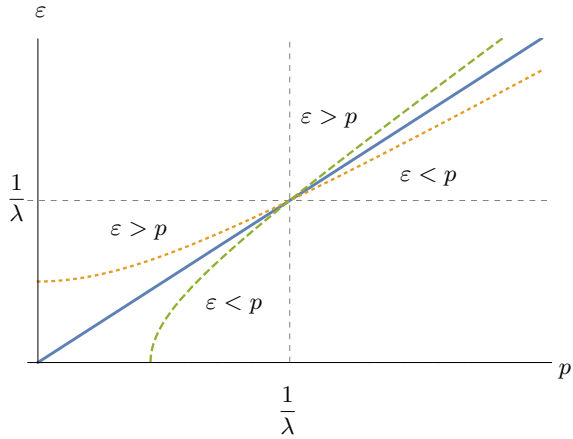

\begin{lpic}[l(7mm),r(4mm),t(4mm),b(7mm)]{RITspacelike_timelike_regions(0.55)}
\lbl[t]{130,0;$p$}
\lbl[b]{63,-17;$\displaystyle \frac{1}{\lambda}$}
\lbl[b]{27,30;$\varepsilon>p$}
\lbl[b]{50,12;$\varepsilon<p$}
\lbl[b]{73,57;$\varepsilon>p$}
\lbl[b]{97,43;$\varepsilon<p$}
\lbl[l]{2,85;$\varepsilon$}
\lbl[l]{-5,40;$\displaystyle \frac{1}{\lambda}$}
\end{lpic}
\caption{\footnotesize Deformed dispersion law plot for bradyons, Eq.\,(\ref{C9}) (dotted line) and for tachyons, Eq.\,(\ref{C9T}) (dashed line).}
\label{crossing}
\end{figure}

\nn Further generalizing our theory we could consider also deformation functions for {\it tachyons} taking a negative right side 
in (\ref{C9}) 
\bb
\frac{\varepsilon^2-p^2}{1-\lambda^2 p^2} = - m^2 
\label{C9T}
\ee
As it appears in Fig.\,(\ref{crossing}), for tachyons the energy-momentum relation is (quasi) symmetrical with respect to the one
holding for bradyons: tachyons are spacelike ($\vare^2 - p^2 < 0$) for $p<1/\lambda$ or timelike ($\vare^2 - p^2 > 0$) 
for $p>1/\lambda$. For $p$ equal to the cut-off tachyons behave as Planckian particles.

\section{More general deformation functions}

\noindent
The present theory can be extended to more general dispersion laws by recourse to a 4x4 matrix deformation function $\widehat{F}(p)$, which in general implies a symmetry breaking between the 4 different spacetime directions:
\bb
\widehat{F}^{\mu\rho}p_\rho \widehat{F}_{\mu\sigma}p^\sigma = m^2
\label{FpFp}
\ee
\nn An important particular case is the diagonal one (see e.g.\,Eq.\,($\ref{Dispersion}$):
\bb 
g^2(p)\varepsilon^2 -f^2(p)p^2 = m^2 
\ee
which corresponds to the following deformation matrix
\bb
\widehat{F} =
\begin {pmatrix}
g(p) & 0 & 0 & 0 \\
0 & f(p) & 0 & 0 \\
0 & 0 & f(p) & 0 \\
0 & 0 & 0 & f(p) \\
\end {pmatrix}
\ee
where $g(p)$ and $f(p)$ are two different momentum functions.
It can be easily proved that a modified LT 
\bb
\begin {pmatrix}
\varepsilon^\prime  \\
p_x^\prime \\
p_y^\prime \\
p_z^\prime \\
\end {pmatrix}
=\Lambda_{\rm mod}
\begin{pmatrix}
\varepsilon  \\
p_x \\
p_y \\
p_z \\
\end {pmatrix}\label{pprimo}
\ee
obeying the invariant dispersion law (\ref{FpFp})
$$
\widehat{F}(p\pr)^{\mu\rho}{p\pr}_\rho \widehat{F}(p\pr)_{\mu\sigma}{p\pr}^\sigma = g^2(p\pr){\varepsilon\pr}^2 -f^2(p\pr){p\pr}^2 = 
$$
$$
= \widehat{F}(p)^{\mu\rho}p_\rho \widehat{F}(p)_{\mu\sigma}p^\sigma = g^2(p)\varepsilon^2 -f^2(p)p^2 = m^2
$$
is given by the following matrix
\[
\Lambda_{\rm mod} = \widehat{F}^{-1}(p\pr)\Lambda\widehat{F}(p)=
\]
\[
= \begin {pmatrix}
g^{-1}(p\pr) & 0 & 0 & 0 \\
0 & f^{-1}(p^\prime) & 0 & 0 \\
0 & 0 & f^{-1}(p^\prime) & 0 \\
0 & 0 & 0 & f^{-1}(p^\prime)\\
\end {pmatrix}\times
$$
$$
\times\begin {pmatrix}
\gamma & -\gamma V & 0 & 0\\
-\gamma V & \gamma & 0 & 0\\
0 & 0 & 1 & 0\\
0 & 0 & 0 & 1\\
\end {pmatrix}\times
\begin {pmatrix}
g(p) & 0 & 0 & 0 \\
0 & f(p) & 0 & 0\\
0 & 0 & f(p) & 0\\
0 & 0 & 0 & f(p)\\
\end {pmatrix}=
\]
\\
\bb
\footnotesize
\begin {pmatrix}
\gamma g^{-1}(p\pr)g(p) & -\gamma Vg^{-1}(p\pr)f(p)  & 0 & 0\\
-\gamma Vf^{-1}(p^\prime)g(p)& \gamma f^{-1}(p^\prime)f(p) & 0 & 0\\
0 & 0 & f^{-1}(p\pr)f(p) & 0\\
0 & 0 & 0 & f^{-1}(p\pr)f(p)\\
\end {pmatrix}
\label{Lambda_def_matrix}
\ee
\\
Just as an example, let us deduce the explicit expression of the MLT when $g(p)=1$ and $f(p)$ is given by ($\ref{Fn}$) for $n=2$, namely
\bb
\displaystyle \varepsilon^2 - \frac{p^2}{1-\lambda^2 p^2}=m^2
\ee
On applying Eq.\,($\ref{Lambda_def_matrix}$), after some algebraic manipulation we get:
\begin{eqnarray}
&&\left\{
\begin{array}{ll}
\displaystyle
\varepsilon^\prime=\gamma\left[\varepsilon-\displaystyle\frac{p_xV}{(1-\lambda^2 p^2)^\frac{1}{2}}\right]\\
p_x\pr=B\gamma\left[p_x-V\varepsilon(1-\lambda^2p^2)^\frac{1}{2}\right]\\
p_y\pr=Bp_y\\
p_z\pr=Bp_z\\
\end{array}
\right.
\end{eqnarray}
where
$$
B\equiv\left\{1-\lambda^2p^2+\lambda^2[\gamma^2[p_x-V\varepsilon(1-\lambda^2p^2)^\frac{1}{2}]^2+p_y^2+p_z^2]\right\}^{-\frac{1}{2}}
$$

\

\nn As a matter of fact, in the present case, and in general for a matrix deformation function, a unique deformed Lorentz factor $A\gamma$ 
for any 4-momentum component cannot be defined as made in the previous section for a non-matrix $C$-number deformation function.

\section{Deformed Lorentz group and spacetime operators\label{DLg}}

\nn In order to get the deformed Lorentz algebra, just taking into account the arising of noncommuting coordinates, 
we adopt a technique often used \cite{Jellal} in noncommutative quantum mechanics as an alternative to the Moyal product: namely, the introduction of suitable auxiliary commuting coordinates. Let us point out that in the present theoretical context the auxiliary coordinates turn out to be non only inter-commuting as usual but, at the same time, relativistically covariant and belonging to the ordinary undeformed Lorentz-Poincaré group. Thence the two properties, commutativity and covariance, are simultaneously owned by the new coordinates (which for this reason can be called ``canonical'').

\nn In particular, we can follow the theoretical approach proposed in \cite{Heuson} and introduce a pair of auxiliary undeformed (i.e. relativistically covariant) canonical variables \textbraceleft ${\pi , \xi}$\textbraceright \, 
defined as 
\bb
\pi^\mu \equiv Fp^\mu \qquad  \qquad \xi^\mu \equiv F^{-1}x^\mu
\label{pi_coordinate}
\ee
Because of (\ref{generalMDL}) we have $\pi^\mu \pi_\mu = m^2$ and $\pi^2$ is a relativistic time-like Casimir invariant, 
as expected since the new variables transform according to the ordinary relativity.

\nn Covariant canonical variables $\pi$ and $\xi$ obey standard rules of quantum mechanics in the $\pi$-space
\begin{eqnarray}
&&
\displaystyle\xi_\mu = i {\frac{\pa }{\pa \pi^\mu}} \label{A22}\\
&& \displaystyle [\pi_\mu, \pi_\nu] = [\xi_\mu, \xi_\nu] = 0
\label{canonical_coordinates} \label{B2}
\label{C2}
\end{eqnarray}
By contrast, in the $p$-space we still have $p$-commutativity
\bb
[p_\mu, p_\nu] = 0
\ee
but, as we are going to see, the spacetime coordinates do not commute anymore
\bb
[x_\mu, x_\nu] \neq 0 
\label{noncomm}
\ee
\nn Indeed, by exploiting the well-known chain rule
$$
\displaystyle {\frac{\pa }{\pa \pi^\mu}}
= {\frac{\pa p^\nu}{\pa \pi^\mu}} {\frac{\pa }{\pa p^\nu}}
$$
after some algebra \cite{Heuson} we obtain the explicit form of the position operator:
\bb
x_\mu = F \xi_\mu = F\,i\frac{\pa }{\pa \pi^\mu} = i \left({\frac{\pa}{\pa p^\mu}} - b_\mu\hat{\ps}\right)
\label{x_hat}
\ee
where $b_\mu$ is defined as follows
\bb \displaystyle b_{\mu} \equiv {\frac{F_{\mu}}{F+\ps F}}
\label{b_mu_definition}
\qquad \quad \displaystyle F_{\mu} \equiv {\frac{\pa F}{\pa p^\mu}}
\ee
and
\bb \displaystyle
\hat{\ps} \equiv p^\lambda \frac{\pa}{\pa p_\lambda}
\ee
can be recognized as a kind of ``scaling/dilation'' operator. Actually, it can be shown [see below, Eq.\,(\ref{B5})]
that operator $x_\mu$ given by (\ref{x_hat})  obeys Eq.\,(\ref{noncomm}).
\\

\nn The 4-rotator $\widetilde{M}^{\mu \nu}$ in the \textbraceleft${\pi,\xi}$\textbraceright-space is defined as usual\footnote{Tensor
$\widetilde{M}^{\mu \nu}$ explicitly writes as
$$
\widetilde{M}^{\mu \nu}= i\frac{\pa}{\pa\pi_\mu}\pi^\nu - i\frac{\pa}{\pa\pi_\nu}\pi^\mu 
$$
Notice that, as is easily seen by applying $\dis i\frac{\pa}{\pa\pi_\mu}\pi^\nu - i\frac{\pa}{\pa\pi_\nu}\pi^\mu$ to a generic quantum 
state $\psi(\pi)$, definition (\ref{Wtilde}) is equivalent to the following one
$$
\widetilde{M}^{\mu \nu} =  \pi^\nu\xi^\mu - \pi^\mu\xi^\nu = i\pi^\nu\frac{\pa}{\pa\pi_\mu} - i\pi^\mu\frac{\pa}{\pa\pi_\nu}
$$
}
\bb
\widetilde{M}^{\mu \nu} =  \xi^\mu\pi^\nu - \xi^\nu \pi^\mu 
\label{Wtilde}
\ee
Actually, because of (\ref{pi_coordinate}), $\widetilde{M}^{\mu \nu}$ and the corresponding operator $\widetilde{M}^{\mu \nu}$ 
defined in the \textbraceleft ${p,x}$\textbraceright-space do coincide
\bb
\displaystyle
\widetilde{M}^{\mu\nu} = \xi^\mu\pi^\nu - \xi^\nu \pi^\mu = M^{\mu\nu} = x^\mu p^\nu - x^\nu p^\mu
\label{defAM}
\ee

\noindent Let us remark that the spin vector modulus turns out to be a spacelike Casimir invariant in both physical and canonical 
representations. In fact, because of (\ref{pi_coordinate}) and (\ref{low_energy_limit}), the Pauli-Lubanski 4-vectors
\begin{eqnarray}
&& \hspace*{-1cm} \displaystyle
W^\mu \equiv \frac{1}{2m} {\varepsilon^{\mu}} _{\nu \rho \sigma} p^\sigma M^{\nu
\rho} = \frac{1}{m} (\sbf \cdot \imp; p^0\sbf -
\imp\times\kbf\,)\label{A1}\\
&& \hspace*{-1cm}  \displaystyle
\widetilde{W}^\mu \equiv \frac{1}{2m} {\varepsilon^{\mu}}_{\nu \rho \sigma} \pi^\sigma {\widetilde{M}^{\nu \rho}}= \frac{1}{m} (\sbf \cdot \pibf; \pi ^0\sbf -
\pibf \times\kbf\,) \label{B1}
\label{C1}
\end{eqnarray}
lead to the same spinorial representation 
\bb 
\displaystyle \widetilde{W}^\mu\widetilde{W}_{\mu} = W^\mu W_{\mu} = - \sbf_\star^2
\ee
$\sbf_\star$ being the spin angular momentum in the center-of-mass ($\imp=0$) reference frame.

\nn With the aid of spacetime operator (\ref{x_hat}) we are now able to derive the new algebra of the Poincaré group in the present 
DSR framework. Actually, by exploiting Eq.\,(\ref{x_hat}), the ordinary angular momentum tensor results to be
\bb 
M_{\mu\nu} =  i \left(p_\nu {\frac{\pa}{\pa p^\mu}} -
p_\mu {\frac{\pa}{\pa p^\nu}} + B_{\mu\nu}\hat{\ps} \right)
\label{M_munu} \ee where \bb B_{\mu\nu} \equiv p_\mu b_\nu - p_\nu
b_\mu 
\ee
For the angular momentum tensor the following equations hold
\begin{eqnarray}
&&
[M_{\mu\nu},p^2] \neq 0
\label{comm_M_p2}\\
&&
[ M_{\mu\nu},\pi^2]=[M_{\mu\nu},F^2 p^2]=0 \label{B4}
\label{C4}
\end{eqnarray}
which are related to the different covariance properties of the non-scalar quantity $p^2$ ($\neq m^2$) and 
of the scalar quantity $\pi^2$ ($= m^2$).\\

\nn The other basic commutators are
\begin{eqnarray}
&&
[p_\mu,p_\nu]=0 \label{A5}\\
&&
[x_\mu,x_\nu]=i(b_\mu x_\nu - b_\nu x_\mu)\label{B5}\\
&&
[x_\mu,p_\nu]=i(g_{\mu\nu} - b_\mu p_\nu)\label{C5}\\
&&
[M_{\mu\nu},x_\sigma]=i(g_{\mu\sigma}x_\nu - g_{\nu\sigma}x_\mu -
B_{\mu\nu} x_\sigma) \label{Mx}\label{D5}\\
&&
[M_{\mu\nu},p_\sigma]=i(g_{\mu\sigma}p_\nu - g_{\nu\sigma}p_\mu + B_{\mu\nu} p_\sigma) \label{Mp} \\
&&
[M_{\mu\nu},M_{\rho\sigma}]= i(g_{\mu\rho}M_{\nu\sigma} - g_{\nu\rho}M_{\mu\sigma}
\label{E5} \\
&&
\qquad\qquad\qquad - g_{\mu\sigma}M_{\nu\rho} + g_{\nu\sigma}M_{\mu\rho} ) \nonumber
\label{G5}
\end{eqnarray}
From the above equations the commutators of the boost operators $N_i (\equiv M_{0i}$ for $i=1,2,3$) 
with $x$ and $p$ acquire new $\lambda$-terms
\begin{eqnarray}
&&
[N_i,p_\sigma]=i(g_{i\sigma}\varepsilon - g_{0\sigma}p_i + B_{0i} p_\sigma)
\label{boost_op_commutator}\\
&&
[N_i,x_\sigma]=i(g_{i\sigma}t - g_{0\sigma}x_i - B_{0i} x_\sigma)\label{A7}
\label{Nx}
\end{eqnarray}
As expected, the ordinary undeformed algebra is recovered in Eqs.\,($\ref{comm_M_p2}-\ref{Nx}$) 
taking $\lambda \longrightarrow 0$. 
Let us also remark that a noncommutative geometry carries new properties and phenomena upon both mathematical and 
physical points of view. In particular, we can expect that Heisenberg Principle applies also to the coordinate uncertainty:
$$
\delta x_\mu \delta x_\nu \neq 0
$$
On applying the new algebra to deformed dispersion law (\ref{Fn}) we have
\bb
b_\mu = \lambda^np^{n-2}p_ig_{\mu i} \qquad\quad B_{0i} =  -\lambda^np^{n-2}\varepsilon p_i
\ee
As a consequence, Eqs.\,(\ref{boost_op_commutator}) and (\ref{Nx}) become
\begin{eqnarray}
&&
\fbox{$[N_i,\varepsilon] = - i(1 + \lambda^np^{n-2}\varepsilon^2)p_i$}
\label{Nvare}
\\
&&
\fbox{$[N_i,p_j] = i (g_{ij} - \lambda^np^{n-2}p_i p_j)\vare$}
\label{Np}
\end{eqnarray} 
and
\begin{eqnarray}
&&
\fbox{$[N_i,t] = - i (x_i - \lambda^np^{n-2}\varepsilon p_i t)\label{A6}$}
\\
&&
\fbox{$[N_i,x_j] = i (g_{ij}t + \lambda^np^{n-2}\varepsilon p_i x_j)\label{B6}$}
\label{C6}
\end{eqnarray} 
Since $[N_i,p_j]$ and $[N_i,x_j]$ result nonvanishing even for $i\neq j$, we expect that a Lorentz boost does affect even transverse momenta, coordinates and lengths.

\nn The other non-vanishing brackets are:
\begin{eqnarray}
&&
[x_i,x_j] = i \lambda^np^{n-2}(p_i x_j - p_j x_i)\\
&&
[t,x_i] = - i \lambda^np^{n-2}p_i t\\
&&
[t,\varepsilon] = i\\
&&
[x_i,p_j] = i (g_{ij} - \lambda^np^{n-2}p_i p_j)
\end{eqnarray}
Finally, let us stress that the time operator \mbox{$t=\dis i\pa/\pa\vare$} given by (\ref{x_hat}) remains ``undeformed'', 
whilst a non-canonical term of the order of $\lambda^2$, at the origin of the underlying noncommutative geometry, 
appears in the position operator 
\bb
\xbf = - i {\frac{\pa}{\pa \imp}} + i\lambda^np^{n-2}\imp\,\hat{\ps}     
\label{x_operator}
\ee
In the special case (\ref{C9}) the above expression reduces to
\bb
\xbf = - i{\frac{\pa}{\pa \imp}} + i\lambda^2\imp\,\hat{\ps}    
\label{imp_operator}
\ee%

\section{Conclusions}

\noindent Deformed SR is a very promising semiclassical (or semiquantal) approach to a final 
theory of quantum spacetime and quantum gravity, trying to address a cardinal problem of contemporary physics,
i.e. the discrepancy of predictions from general relativity and from quantum mechanics and standard model. 
The key tool adopted in any DSR model is the introduction of a very large momentum cut-off and, correspondingly, 
of a  very small elementary spatial length $\lambda$. These fundamental scales are at the same time just the 
scales of the Lorentz symmetry violation involved by any DSR theoretical model. Actually, the momentum cut-off 
results in avoiding infinities occurring for some physical quantities in relativistic quantum field theory at 
very high energies and in general relativity at very short distances. In DSR even the quantum vacuum total energy
or the zero point energy result to be very large but finite quantities (for instance, it can be shown that in our
theory the ZPF energy is finite and proportional to $\lambda^{-4}$). 

\

\noindent The present one is only a preliminary seminal work, based on an essentially phenomenological approach,
which will be continued in subsequent studies and researches on possible observable consequences and applications 
in astrophysics, particle physics and cosmology. In particular, DSR deviations from SR predictions are often considered
 in literature for particle decays and scattering. Also on a mere theoretical point of view, DSR models as the present one 
entail some interesting puzzling problems as, e.g., apparent energy nonconservation (\cite{Agostini}) still to 
be investigated, and appear as promising approaches to SR and general relativity reformulations in the phase space (rather
than in the ordinary configuration space) which might be needed for a proper quantum theory of gravitation.

\

\noindent In this paper we have focused on the derivation of explicit modified Lorentz transformations 
for a large class of ``deformations" of SR due to the introduction of a suitable family of 
non-covariant momentum-energy dispersion laws. Therefore our approach violates the relativistic covariance, 
even if the Lorentz symmetry breaking is effective only at Planckian energies and momenta, while in the low
energy sector ordinary SR is recovered.
Moreover, as is well-known, non-covariant dispersion laws entail not only a deformed Lorentz group, but 
also a noncommutative spacetime geometry. In DSR we find a noncommutative geometry also for curved 
spacetimes where it is often recognized as a kind of a ``quantum Riemann geometry'' \cite{Ruegg}. 

\nn We have first derived the explicit form of MLTs in correspondence of an important set of deformation
functions. Besides a novel and straight deduction technique, the MLTs for any $n$ are yet unpublished and 
look as a direct extension or generalization of the ordinary LTs.
Then we have analyzed the behavior of the deformed Lorentz factor as a function of the particle 
momentum and of the boost speed: for quasi-luminal boosts $V\sim c$, applied to the quiet frame, this factor 
does not diverge but approaches a large but finite value $1/\lambda m$ for any $n$. We expect that this result 
should affect the usual relativistic phenomenology as length contraction, time dilation, mass increase for very 
high speed boosts $V>1/\sqrt{1+m^2\lambda^2}$. 

\

\noindent Noticeably, if we would extend our approach beyond the momentum cut-off, under a generic 
(ordinary, i.e. subluminal) boost timelike particles ($\vare^2 - p^2 > 0$), luxons ($\vare^2 - p^2 = 0$), and  spacelike
particles ($\vare^2 - p^2 < 0$) would remain separate away each other as in ordinary relativity, so that the sign of the 
effective (variable) mass squared would not vary under deformed Lorentz transformations. Moreover, a special class of  
bradyons, which we have named ``Planckian particles'' in that endowed with Planck energy and momentum, might be classified 
as a further fourth particle type. In fact these particles, despite of being massive, travel with light speed as photons 
but remain Planckian particles under any boost and with respect to any observer (while photon 4-momentum depends 
on the given reference frame).

\

\nn Finally we have extended our theory to the case of matrix deformation functions, with an explicit
application to the case of diagonal matrices.  In so doing we have generalized our theoretical approach to deformed 
dispersion laws containing momentum-energy mixed terms, obtaining the modified Lorentz transformations also for 
special dispersion laws similar to the ones involved by $k$-Minkowski approaches to DSR.

\noindent In the last Section we have also reviewed the main algebraic properties of the deformed spacetime 
translation and rotation generators in correspondence of generic dispersion law and deformation function.

\end{document}